\documentclass[iop,apj,numberedappendix,appendixfloats]{emulateapj}

\usepackage{amsmath}
\usepackage{multirow}
\usepackage{graphicx,float}
\usepackage{lineno}

\bibpunct[, ]{(}{)}{;}{a}{}{,}

\newcommand{\Omegam}{\ensuremath{\Omega_{\mathrm{m}}}}
\newcommand{\OmegaL}{\ensuremath{\Omega_{\Lambda}}}

\newcommand{\hgvol}{h^{-3}{\mathrm{Gpc}}^{3}}

\newcommand{\hmpc}{h^{-1}\mathrm{Mpc}}

\newcommand{\lcdm}{\ensuremath{\Lambda\mathrm{CDM}}}

\begin{document}

\title{The Alcock-Paczy\'nski Test Using Cosmic Voids in the Final BOSS DR12 Galaxy }
\title{Cosmic Voids in the SDSS DR12 BOSS Galaxy Sample: The Alcock-Paczy\'nski Test}

\author{
  Qingqing~Mao\altaffilmark{1},
  Andreas~A.~Berlind\altaffilmark{1,2},
  Robert~J.~Scherrer\altaffilmark{1},
  Mark~C.~Neyrinck\altaffilmark{3},
  Rom\'an~Scoccimarro\altaffilmark{4},
  Jeremy~L.~Tinker\altaffilmark{4},
  Cameron~K.~McBride\altaffilmark{5},
  Donald~P.~Schneider\altaffilmark{6}
}
\altaffiltext{1}{Department of Physics and Astronomy, Vanderbilt University, Nashville, TN 37235, USA}
\altaffiltext{2}{a.berlind@vanderbilt.edu}
\altaffiltext{3}{Department of Physics and Astronomy, The Johns Hopkins University, Baltimore, MD 21218, USA}
\altaffiltext{4}{Center for Cosmology and Particle Physics \& Department of Physics, New York University, New York, NY 10003, USA}
\altaffiltext{5}{Harvard-Smithsonian Center for Astrophysics, Cambridge, MA 02138, USA}
\altaffiltext{6}{Institute for Gravitation and the Cosmos, Department of Astronomy and Astrophysics, The Pennsylvania State University, University Park, PA 16802, USA}


\begin{abstract}
We apply the Alcock-Paczy\'nski (AP) test to the stacked voids identified using the large-scale structure galaxy catalog from the Baryon Oscillation Spectroscopic Survey (BOSS). This galaxy catalog is part of the Sloan Digital Sky Survey Data Release 12 and is the final catalog of SDSS-III. We also use 1000 mock galaxy catalogs that match the geometry, density, and clustering properties of the BOSS sample in order to characterize the statistical uncertainties of our measurements and take into account systematic errors such as redshift space distortions. For both BOSS data and mock catalogs, we use the ZOBOV algorithm to identify voids, we stack together all voids with effective radii of $30-100\hmpc$ in the redshift range $0.43-0.7$, and we accurately measure the shape of the stacked voids. Our tests with the mock catalogs show that we measure the stacked void ellipticity with a statistical precision of 2.6\%. The stacked voids in redshift space are slightly squashed along the line of sight, which is consistent with previous studies. We repeat this measurement of stacked void shape in the BOSS data assuming several values of $\Omegam$ within the flat $\lcdm$ model, and we compare to the mock catalogs in redshift space in order to perform the AP test. We obtain a constraint of $\Omegam = 0.38^{+0.18}_{-0.15}$ at the $68\%$ confidence level from the AP test. We discuss the various sources of statistical and systematic noise that affect the constraining power of this method. In particular, we find that the measured ellipticity of stacked voids scales more weakly with cosmology than the standard AP prediction, leading to significantly weaker constraints. We discuss how AP constraints will improve in future surveys with larger volumes and densities.
\end{abstract}


\section{Introduction}
Understanding the nature of dark energy, which drives cosmic acceleration \citep{Riess:1998, Perlmutter:1999}, is the premier goal of present day cosmology. To constrain the wide variety of dark energy models, it is important to apply several complementary statistical methods to available observational data \citep{Weinberg:2013}. A classical method for probing the expansion history of the universe was proposed by \citet{Alcock:1979}. The Alcock-Paczy\'nski test (AP test) is a purely geometric test that examines the ratio of the observed angular size to radial size of objects that are known to be intrinsically isotropic. Most applications of the AP test have focused on measuring the anisotropic clustering of galaxies using the correlation function or power spectrum. Another approach is to measure the symmetry properties of close galaxy pairs \citep{Marinoni:2010}. Unfortunately, this method is seriously affected by dynamics at small scales \citep{Jennings:2012}. 

First proposed by \citet{Ryden:1995} and extensively discussed by \citet{Lavaux:2012}, cosmic voids provide an attractive alternative for applying the AP test. Voids are large underdense regions present in the large scale structure of the Universe. Ever since their discovery more than 30 years ago \citep{Gregory:1978, Kirshner:1981, Lapparent:1986}, voids have been recognized as interesting laboratories for studying cosmology. The low-density nature of voids places them in the quasi-linear regime, thus it may be easier to model systematics such as redshift-space distortion (RSD) effects. Although the shapes of individual voids can be very irregular, the AP test can be applied to stacked voids to significantly reduce this "shape noise". To successfully apply the AP test to voids, a large volume spectroscopic galaxy redshift survey is essential. \citet{Lavaux:2012} reported that the constraining power of this method could rival that of the Baryon Acoustic Oscillations \citep[BAO;][]{Eisenstein:2005} method when applied to the completed Baryon Oscillation Spectroscopic Survey \citep[BOSS;][]{Dawson:2013} and surpass it for even larger future surveys.

To date, two studies have applied the AP test on stacked cosmic voids. \citet{Sutter:2012b} used galaxy catalogs from the Sloan Digital Sky Survey \citep[SDSS;][]{York:2000} Data Release 7 \citep[DR7;][]{Abazajian:2009}, but the data did not have sufficient signal-to-noise ratio to detect the AP effect. \citet{Sutter:2014a} examined the SDSS Data Release 10 \citep[DR10;][]{Ahn:2014} and definitively detected the AP signal. In this work, we use the galaxy catalogs from the SDSS Data Release 12 \citep[DR12;][]{Alam:2015}, which is the final data release of BOSS and the largest available spectroscopic survey at the present time. We identify voids using the ZOBOV void-finding algorithm \citep{Neyrinck:2008}, and measure the shape of the stacked voids. In addition, we use 1000 galaxy mock catalogs to characterize the uncertainties and correct for the systematics such as RSD effects. These void catalogs are described in detail in \citet{Mao:2016}.

This paper is organized as follows. We first briefly introduce the AP test in \S\ref{s:ap_test}. In \S\ref{s:data}, we describe the galaxy and mock catalogs used in this study. In \S\ref{s:finding_voids}, we describe the method and the steps we take to identify the voids. We then discuss how to stack the voids in \S\ref{s:stacking} and how to accurately measure the shape of stacked voids in \S\ref{s:shape}. In \S\ref{s:results} we use our AP test results to derive cosmological constraints. Finally, we discuss our results in \S\ref{s:discussion}. 

\section{Alcock-Paczy\'nski Test}
\label{s:ap_test}
Consider an intrinsically spherical object at redshift $z$ with an extension of $\Delta z$ in the line-of-sight direction and $\Delta \theta$ across the sky. In comoving coordinates, the object has size $\Delta l$ in the line-of-sight direction and $\Delta r$ in the transverse direction. Then $\Delta l$ is related to $\Delta z$ by
\begin{equation}
	\Delta l = \frac{c}{H(z)} \Delta z, 
	\label{eq:deltal}
\end{equation}
where $H$ is the Hubble parameter. In a flat $\lcdm$ universe, 
\begin{equation}
	H(z) = H_0 \sqrt{\Omegam (1+z)^3 + \OmegaL},
\end{equation}
where $H_0$ is the present value of the Hubble parameter, $\Omegam$ is the present value of the matter density parameter, and $\OmegaL$ is the present value of the dark energy density. The transverse comoving size $\Delta r$ is related to $\Delta \theta$ by 
\begin{equation}
	\Delta r = D_M(z) \Delta \theta = (1+z) D_A(z) \Delta \theta, 
	\label{eq:deltar}
\end{equation} 
where $D_M$ is the transverse comoving distance and $D_A$ is the angular diameter distance \citep{Hogg:1999}. In a flat universe, the transverse comoving distance is equal to the line-of-sight comoving distance,  
\begin{equation}
	D_M(z) = D_C(z) = c \int_0^z \! \frac{\mathrm{d} z'}{H(z')}. 
\end{equation}
Since $\Delta l$ and $\Delta r$ are equal for a spherical object, combining equation~\ref{eq:deltal} and \ref{eq:deltar} yields the ratio 
\begin{equation}
	\frac{\Delta z}{z \Delta \theta} = \frac{(1+z)}{cz} D_A(z)H(z). 
\end{equation}
This is the original form of the AP test, and one can directly compare the observable quantity on the left hand side to the prediction from a cosmological model on the right hand side.

Another way to view the AP test is that one will only recover the spherical symmetry of the object when assuming the true cosmology. If we convert redshift to comoving distance by assuming a cosmological model, we can measure the ratio
\begin{equation}
	e(z) = \frac{\Delta l'}{\Delta r'} = \frac{\Delta l' / \Delta r'}{\Delta l / \Delta r} = \frac{D_A(z)H(z)}{D_A'(z)H'(z)}, 
	\label{eq:ratio}
\end{equation}
where primes indicate quantities calculated using the assumed cosmology and unprimed values are calculated using the true cosmology. We can then measure the ratio $e(z)$ with a set of different assumed cosmologies and find the cosmology that yields $e(z)=1$.

In this study, we assume a flat $\lcdm$ model with a cosmological constant. In this case the AP test only depends on a single parameter, $\Omegam$. We adopt a set of fiducial cosmologies with different values of $\Omegam$, and for each cosmology we convert redshifts to comoving distances, identify voids in both the BOSS galaxy catalog and the mock galaxy catalogs, and measure the ratio $e(z)$ of the stacked voids. 

\section{Data and mocks}
\label{s:data}
The galaxy sample used in this study is from the Baryon Oscillation Spectroscopic Survey (BOSS; \citealt{Dawson:2013}), which is part of the third generation of the Sloan Digital Sky Survey (SDSS-III; \citealt{Eisenstein:2011}). BOSS made use of the dedicated SDSS telescope \citep{Gunn:2006}, multi-object spectrograph \citep{Smee:2013}, and software pipeline \citep{Bolton:2012}, to obtain the spectra of over 1.37 million galaxies over two large contiguous regions of sky in the Northern and Southern Galactic Caps, covering over 10,000 $\mathrm{deg}^2$ in total. DR12 is the final data release of SDSS-III and contains all five years of BOSS data. 

BOSS galaxies were uniformly targeted in two samples, a relatively low-redshift sample with $z < 0.45$ (LOWZ) and a high-redshift sample with approximately $0.4 < z < 0.7$ (CMASS). A full description for the targeting criteria can be found in \citet{Dawson:2013}. We only include the CMASS sample in this analysis using redshift limits of $0.43 < z < 0.7$. The median redshift of the sample is 0.57 and it has a total volume of $3.7\hgvol$. The CMASS sample is not volume-limited and the number density of galaxies depends on redshift. Whenever we need to compare a local density to the mean density of the sample, we always use the mean density at that redshift, $\bar{n}(z)$. 

The large-scale structure (LSS) galaxy catalog is produced by the BOSS collaboration as a value-added catalog. The catalog includes weights to correct for the effects of redshift failures and fiber collisions. In addition, there are weights to account for the systematic relationships between the number density of observed galaxies and stellar density and seeing. These weights are described in detail in \citet{Reid:2016}. The LSS catalog uses the MANGLE software \citep{Swanson:2008} to account for the survey geometry and the angular completeness. For each distinct region, we up-weight all the galaxies in the region according to its completeness to correct for the angular selection function.

In this study we use a set of 1000 mock galaxy catalogs to estimate statistical uncertainties and study systematics. The mock catalogs were generated using the ``quick particle mesh'' (QPM) methodology described by \citet{White:2014}. These QPM mocks were based on a set of rapid but low-resolution particle mesh simulations that accurately reproduce the large-scale dark matter density field. Each simulation contained $1280^3$ particles in a box of side length 2,560 $\hmpc$. The adopted cosmological parameter values were $\Omegam = 0.29$, $h = 0.7$, $n_s = 0.97$ and $\sigma_8 = 0.8$. Mock halos were selected based on the local density of each particle. These halos were then populated using the halo occupation distribution \citep[HOD;][]{Berlind:2002} method to create galaxy mocks. The HOD was chosen such that the clustering amplitude of mock galaxies matches that of BOSS CMASS galaxies. The survey masks were applied to give the mocks the same survey geometry as the data, and the mocks were randomly down-sampled to have the same angular sky completeness and the same radial density distribution $\bar{n}(z)$ as the data. Finally, redshift space distortions were included based on the velocity of the simulation particles. 

\section{Finding voids}
\label{s:finding_voids}
We use the ZOBOV algorithm to identify voids in the BOSS data and QPM mock catalogs. ZOBOV first applies Voronoi tessellation to assign a Voronoi cell and produce a density estimate for each galaxy. The algorithm then uses the watershed transform to group neighboring Voronoi cells into zones and eventually subvoids and voids. ZOBOV also provides the statistical significance of a void, i.e., the probability that it could arise from a random distribution of points. One of the advantages of ZOBOV is that it does not make any assumptions about void shape, thus allowing us to explore the natural shapes of voids. The detailed description of the algorithm can be found in \citet{Neyrinck:2008}. 

We prepare the LSS catalog to take into account the survey geometry by placing a high number density of randomly distributed buffer particles around the survey boundaries. The purpose of these buffer particles is to ensure the tessellation process even for galaxies close to the survey boundaries, but these boundary galaxies are not included in the watershed transform step. We apply all the weights immediately after the tessellation stage by directly modifying the corresponding density of each galaxy as $n_i = w_i / V_i$, where $w_i$ is the total weight of the galaxy and $V_i$ is the volume of its Voronoi cell. However, we retain all the adjacency information. This is an easy way to include the systematic weights and angular selection function, which allows the watershed transform to run smoothly with no additional modification. 

In general, ZOBOV is parameter free, but some restrictions can be applied as needed. We set the density threshold parameter to 0.5, which limits ZOBOV during the watershed transform step to only group together zones of density less than half the mean density of the whole sample. We also check the minimum Voronoi density of each void (i.e., the galaxy with the largest Voronoi cell in the void) and require that it is less than half of the mean density $\bar{n}(z)$ at the void center's redshift. Finally, we only include voids with significance higher than $2\sigma$. We run the ZOBOV algorithm directly on the redshift space catalog and we do not attempt to remove the redshift distortions in the data. In the case of the QPM mocks, we find voids both in the real space catalogs and in the redshift space catalogs. 

The void finding procedure is identical to that described in our recent public void catalog release \citep{Mao:2016}; the only difference is that in the public catalog we assumed a fixed fiducial cosmology of $\Omegam=0.3$, whereas in this work for the AP test, we assume a fiducial cosmology of  $\Omegam=0.29$, since that is the cosmology that was used to run the QPM simulations (this difference is negligible). Additionally, in this work we assume a series of different values of $\Omegam$ in order to explore the cosmological constraints. For each cosmology, we convert galaxy redshifts to line-of-sight comoving distances using that cosmology before applying the ZOBOV algorithm. 

\section{Stacking voids}
\label{s:stacking}
The weighted center of each void is defined as the average position of the void galaxies weighted by the inverse density, calculated as
\begin{equation}
	\mathbf{X} = \frac{\sum_i \mathbf{x}_i / n_i}{\sum_i 1 / n_i},
	\label{eq:voidcenter}
\end{equation}
where $\mathbf{x}_i$ is the position of each galaxy in the void and $n_i$ is the corresponding weighted density of the Voronoi cell. The size of a void is defined by its effective radius: 
\begin{equation}
	R_\mathrm{eff} \equiv \left(\frac{3}{4\pi}V\right)^{1/3}, 
	\label{eq:Reff}
\end{equation}
where $V$ is the sum of all the Voronoi volumes in the void. 

\begin{figure}
	\includegraphics[scale=0.45]{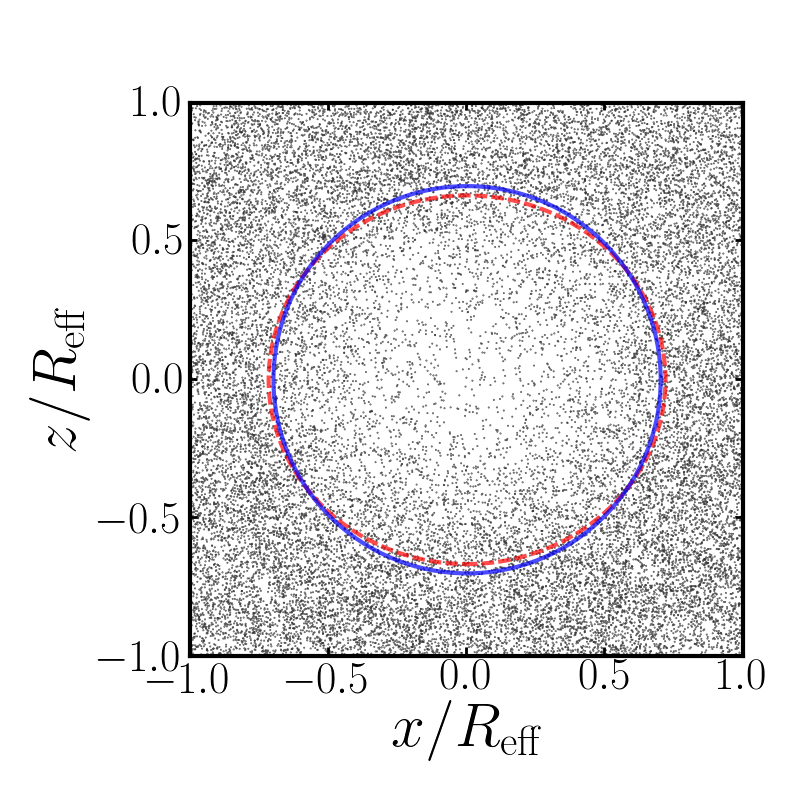}
	\caption{A slice through 717 stacked BOSS CMASS voids with effective radii ranging from 30 $\hmpc$ to 100 $\hmpc$ and assuming $\Omegam=0.29$. Before stacking, each void is rescaled to its effective radius and rotated so that the line-of-sight direction aligns with the z-axis in the figure. Black dots represent all the galaxies around each void center, and not just the void member galaxies. The blue circle has a radius of 0.7, highlighting the approximate region within which we measure the void shape, while the dashed red ellipse has the ellipticity of 0.929 that is measured for this stacked void.}
	\label{fig:stacking_visual_single}
\end{figure}

We stack all voids with effective radii ranging from 30 $\hmpc$ to 100 $\hmpc$, which includes most of the identified voids, on their weighted centers. Each void is first rescaled to its effective radius and rotated so that all voids share a common line-of-sight direction before stacking. The stacking can be done by only using void member galaxies or by using all the galaxies around each void center. Our tests show that both methods produce similar results, but the shape measurements are more stable when using all the galaxies (we discuss this more in the next section). Figure~\ref{fig:stacking_visual_single} presents a slice through the 717 stacked BOSS voids constructed assuming $\Omegam=0.29$. It clearly shows an underdense central region, with a rising density moving towards the outer part of the stacked voids. The density profile of these stacked voids is shown in \citet{Mao:2016}. 

We choose to stack all the voids in the BOSS CMASS sample, rather than in narrow redshift bins as was done by \citet{Sutter:2012b,Sutter:2014a}. We tested this approach and determined that the redshift range of the CMASS sample is not large enough to yield distinguishable results in different redshift bins given the limited number of voids in the sample. \citet{Sutter:2012b} suggested that it is important to avoid using voids too close to the survey boundaries since they might be preferentially aligned relative to the line of sight, and thus bias the result. Using our large set of mock catalogs, we find that we can reliably retrieve the correct shape of stacked voids without worrying about survey boundaries. Our tests show that if we only stack voids that are not near any of the survey boundaries, the shape of the stacked voids is consistent with that made from using all the voids; however, the uncertainties are larger because the number of the voids that are far from any boundaries is limited. We therefore choose to use all voids and we argue that the net effect of survey boundaries must be small. This can happen because the effect is opposite for voids near redshift boundaries (where voids will be preferentially squashed along the line of sight) compared to boundaries on the sky (where voids will be preferentially elongated), thus canceling each other to a large extent.

\section{Shape measurements}
\label{s:shape}
To perform the AP test we must measure an ellipticity for the stacked voids. The void ellipticity can be different from unity due to four causes: (1) the AP effect due to assuming the wrong cosmological model, which is the signal we want; (2) an intrinsic asphericity due to stacking a finite number of non-spherical voids, which we designate "shape noise" and is the dominant source of statistical noise in this measurement; (3) a Poisson or shot noise due to sampling the void with a finite number of galaxies, which is a sub-dominant source of statistical noise; (4) redshift distortions, which are the main source of systematic error. 

Whether the stack includes all galaxies around void centers or only includes void member galaxies, the stacked void does not have a distinct outer boundary. Moreover, the outer region of the void includes high density structures, such as the walls, filaments and clusters that surround voids, which can be strongly affected by redshift-space distortions. We thus prefer to measure the ellipticity using the inner, underdense, part of the voids. The method proposed by \citet{Lavaux:2012} and used by \citet{Sutter:2012b} involves fitting a model of the anisotropy of the stacked void density profile. However, this technique requires a choice of profile shape and is thus somewhat model dependent. We adopt a simpler approach, which is to measure an axis ratio from all the galaxy positions within a chosen volume inside the void. The axis ratio $e$ is given by
\begin{equation}
	e = \sqrt{\frac{2 \sum_i w_i z_i^2}{\sum_i w_i(x_i^2 + y_i^2)}}, 
	\label{eq:shape}
\end{equation}
where $x_i$, $y_i$ and $z_i$ are the galaxy Cartesian coordinates in the stacked void of which $z$ is in the aligned line-of-sight direction, $w_i$ are the galaxy weights, and the summation is taken over all the galaxies within the chosen volume. \citet{Sutter:2014a} also used a similar measurement, although they did not include the galaxy weights.

The axis ratio measured by Equation~\ref{eq:shape} is not a direct measurement of the true ellipticity of the stacked voids because it also depends on the shape of the volume within which it is measured. For example, a truly spherical void (i.e., no stretch due to the AP effect or due to redshift distortions or due to shape or shot noise) will only have a measured axis ratio of unity if the volume is a sphere. If the measuring volume is instead an ellipsoid, then the measured axis ratio will deviate from unity. Conversely, a truly elliptical void will not yield the correct axis ratio if it is measured in a spherical volume. It turns out that a void with a true ellipticity $e_\mathrm{true}$ will only have an axis ratio measurement $e = e_\mathrm{true}$ if the shape of the volume within which the axis ratio is measured is itself an ellipsoid of ellipticity equal to $e_\mathrm{true}$. This result suggests a straightforward method for recovering the true void shape. First, measure the axis ratio $e$ using Equation~\ref{eq:shape} in a series of ellipsoidal volumes of varying ellipticity $e_\mathrm{cut}$. Next, identify the measurement where $e = e_\mathrm{cut}$ and adopt that measurement as the true internal shape of the stacked voids.

\begin{figure}
	\includegraphics[scale=0.45]{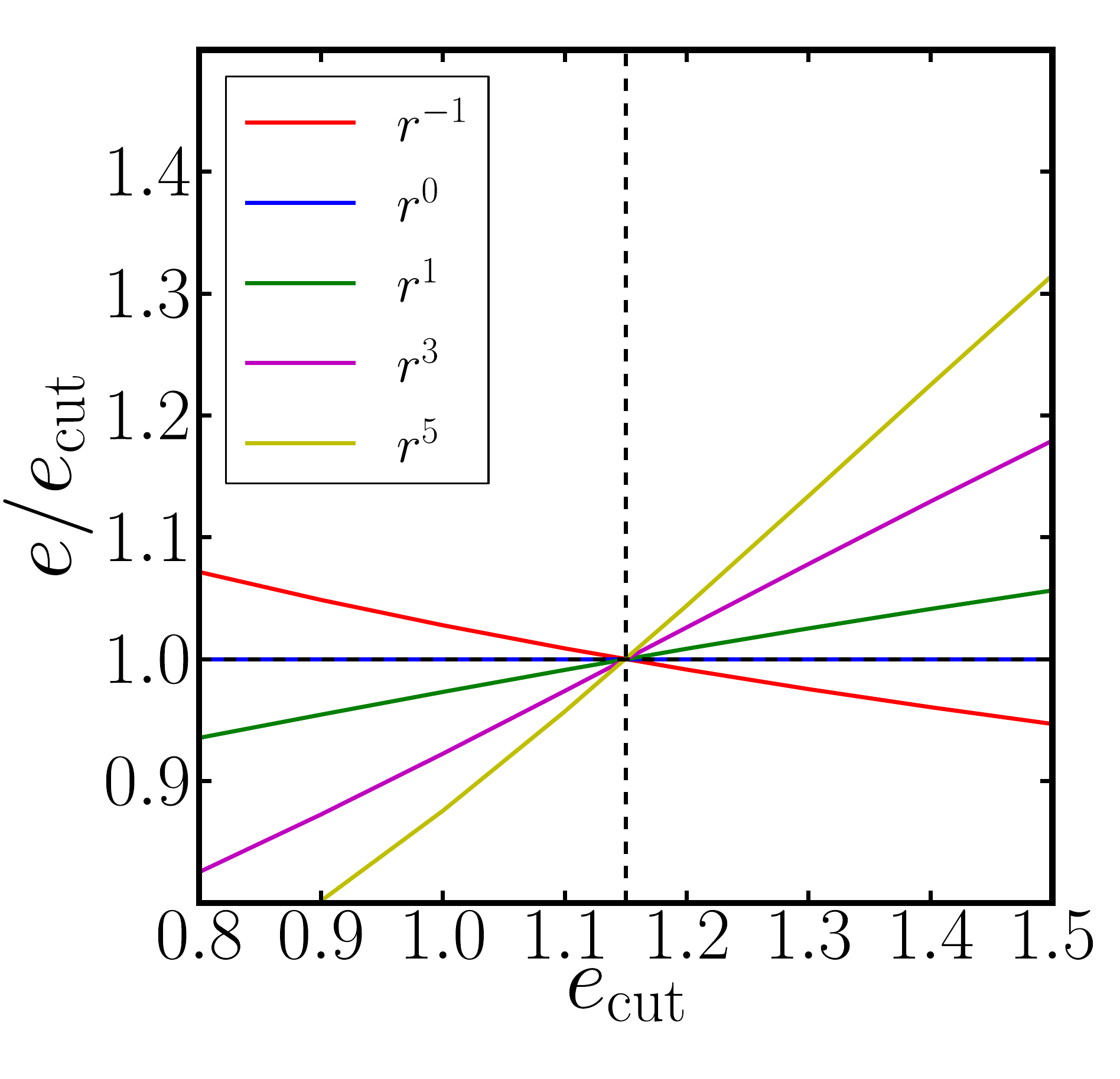}
	\caption{An analytic test of our method for measuring the shape of stacked voids. Each line displays results for a different adopted power-law density profile (as listed in the panel) that is stretched along the line-of-sight to have a true ellipticity of 1.15. For each density profile, we calculate the axis ratio $e$ by integrating the profile over an ellipsoidal volume of varying ellipticity $e_\mathrm{cut}$. We plot the ratio $e / e_\mathrm{cut}$ as a function of $e_\mathrm{cut}$ and show that the ratio is only equal to one for $e_\mathrm{cut} = 1.15$, as indicated by the black vertical dashed line.}
	\label{fig:shape_analytical_test}
\end{figure}

Figure~\ref{fig:shape_analytical_test}, demonstrates via analytic calculation that this method works. In this exercize, we adopt a series of power-law density profiles that are all stretched along the $z$ direction by 15\%, i.e., they have a true ellipticity of 1.15. For each profile, we calculate the axis ratio $e$ by integrating the profile over an ellipsoidal volume of varying ellipticity $e_\mathrm{cut}$, and plot the ratio $e / e_\mathrm{cut}$ as a function of $e_\mathrm{cut}$. The figure shows that for all the adopted density profiles except for the uniform density case, the ratio is only equal to one for $e = e_\mathrm{cut} = 1.15$. Numerical tests using randomly-generated points following the same density profiles give exactly the same results. This method clearly works for any distribution of points exhibiting a density gradient, and should therefore work on stacked voids. 

\begin{figure}
	\includegraphics[scale=0.45]{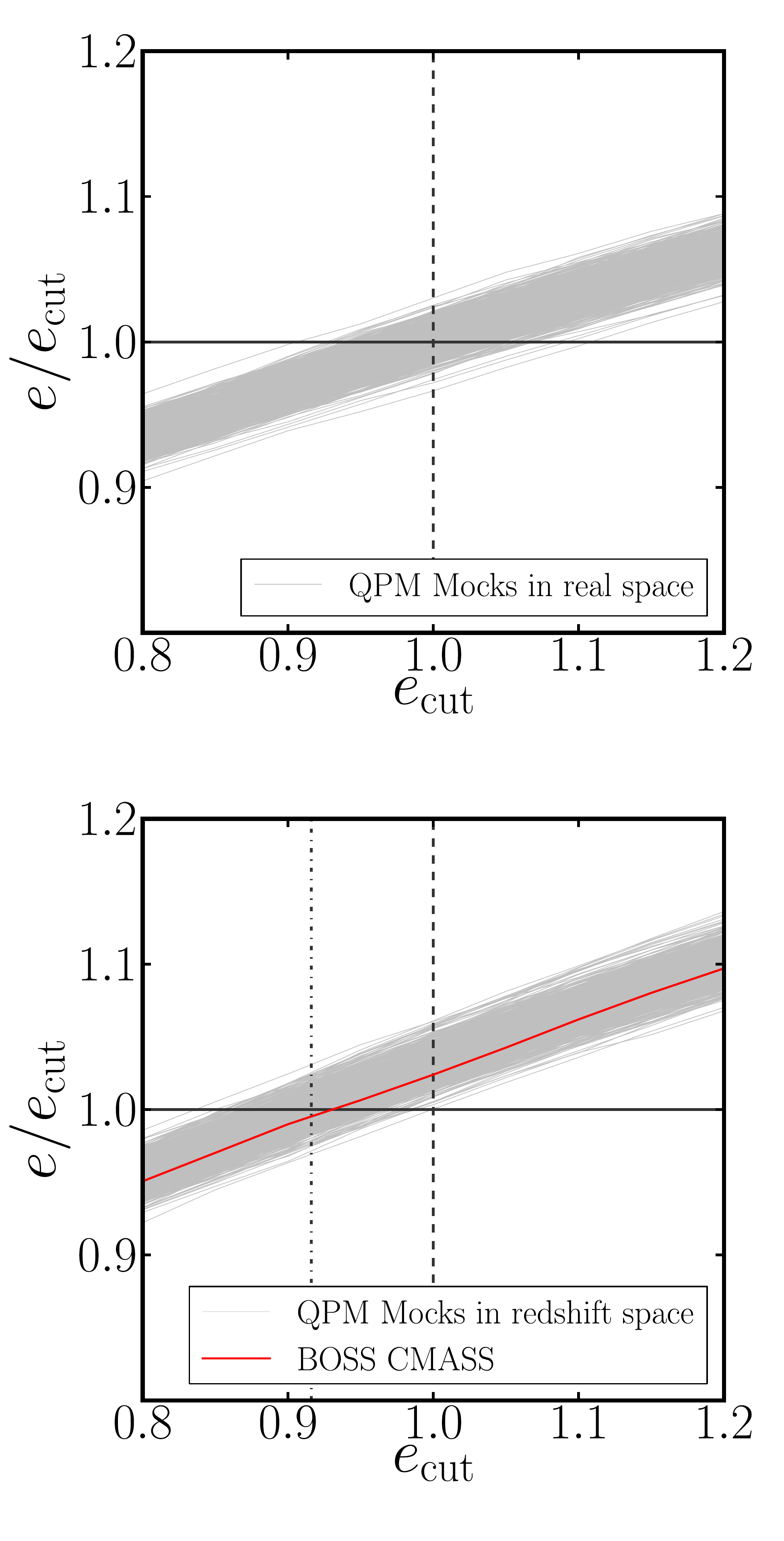}
	\caption{The method of measuring the shapes of stacked voids applied to QPM mock catalogs and BOSS CMASS data, when the fiducial cosmology of $\Omegam=0.29$ is assumed to convert redshifts to comoving distances. As in Fig.~\ref{fig:shape_analytical_test}, we show the ratio of measured axis ratio $e$ to the ellipticity of the ellipsoidal volume $e_\mathrm{cut}$ within which it is measured, as a function of $e_\mathrm{cut}$. The true shape of the stacked voids is then given by the value of $e_\mathrm{cut}$ where this ratio is equal to one (represented by the horizontal black lines). Each grey line shows this result for one of the 1000 QPM mock catalogs in real space (top panel) and redshift space (bottom panel). The red line in the bottom panel shows the result for the BOSS CMASS stacked voids. Vertical dashed lines mark the value $e_\mathrm{cut}=1$ for reference, while the vertical dot-dashed line in the bottom panel shows the mean ellipticity of the stacked voids in the QPM mock catalogs.}
	\label{fig:e_distr}
\end{figure}

\begin{figure}
	\includegraphics[scale=0.45]{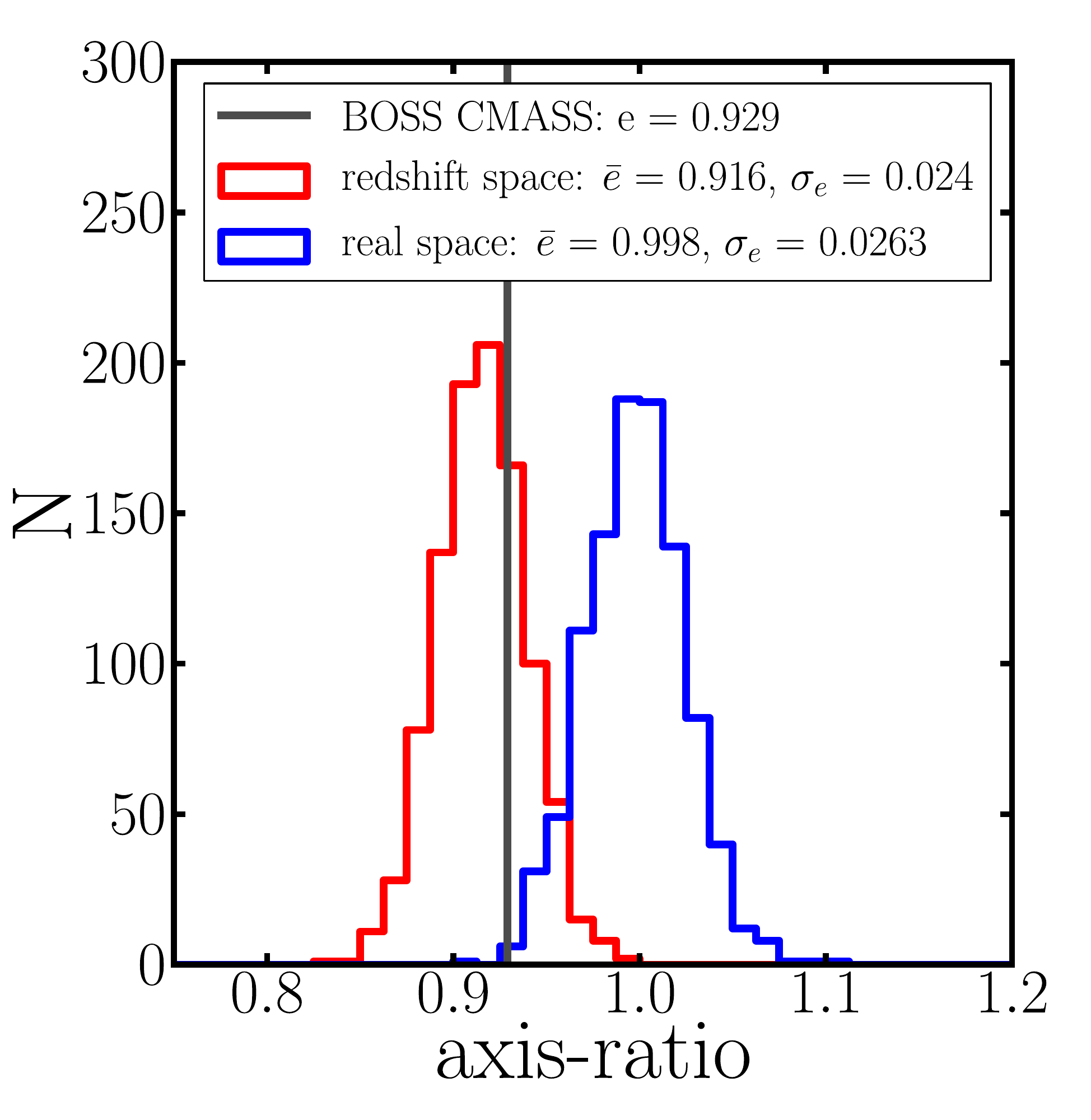}
	\caption{Shape measurements for the BOSS CMASS stacked voids, compared to the 1000 QPM mock catalogs, when the fiducial cosmology of $\Omegam=0.29$ is assumed to convert redshifts to comoving distances. The distribution of mock void shapes is shown in real space (blue histogram) and redshift space (red histogram). Also displayed is the shape measurement from the BOSS data (vertical black line). The BOSS void axis ratio, as well as the mean and standard deviation of the mock void axis ratios, are listed in the panel. Mock voids in real space are spherical, as expected, while in redshift space they are slightly squashed along the line of sight. The shape of the BOSS voids is consistent with the distribution of mock shapes in redshift space.}	
	\label{fig:shape_distr}
\end{figure}

Before applying this method to measure the shapes of our stacked voids, we must make two choices. First, we need to decide whether to include all the galaxies around void centers when stacking, or to only include galaxies that were identified as belonging to the voids. Using all the galaxies has the advantage that our results will not be as sensitive to the uncertainties involved in determining void boundaries. In this case, the only information we use from the void identification is the void center and the void radius. Additionally, using all galaxies around void centers provides more points in the stack, which improves the Poisson noise in the shape measurements. On the other hand, including galaxies exterior to the voids runs the risk of including nearby clusters and other high density regions that can magnify the effect of redshift distortions. We test these two cases on our mock catalogs and determine that, while both methods recover consistent shape measurements, using all galaxies when stacking leads to $\sim30\%$ higher precision than only using void galaxies. This is because using all the galaxies reduces the shape noise produced from voids with highly irregular shapes or from voids with significant errors in centering. We therefore use all the galaxies around void centers in all our analysis.

The second choice is the volume within which we measure the stacked void shape. If we choose too large an ellipsoidal volume, it will likely be affected by the high density regions surrounding voids. If instead we choose too small a volume that only encloses the inner regions of voids, the small number of galaxies will increase the Poisson noise of our shape measurement. We tested the shape measurements using different ellipsoid sizes, and we found that ellipsoids with a volume equal to that of a sphere of radius 0.7 times the effective radius of the stacked voids provide the most stable results and optimize uncertainties. We adopt this volume for all our shape measurements. The blue circle in Figure~\ref{fig:stacking_visual_single} has a radius of 0.7 times the effective void radius, while the red dashed ellipse shows a slightly elongated ellipsoid with the same volume.

We now use the method described above to measure the shapes of stacked voids in the 1000 QPM mock catalogs in real space. We first assume the same fiducial cosmological model, with $\Omegam=0.29$, as was used to run the QPM simulations, and we convert mock galaxy redshifts to comoving coordinates. Next, we run the ZOBOV void finder on each mock catalog and stack all the voids as described above. We then apply an ellipsoidal volume cut on the stacked voids, with a changing ellipticity $e_\mathrm{cut}$ but fixed volume equal to that of a sphere of radius $R=0.7$. In each ellipsoid we measure the axis ratio of the enclosed galaxies, $e$, according to Equation~\ref{eq:shape}. The top panel of Figure~\ref{fig:e_distr} displays the resulting measurements of $e/e_\mathrm{cut}$ as a function of $e_\mathrm{cut}$. Each gray line represents a measurement from one of the 1000 mocks. We do a spline fit to each gray line to determine where the measured $e$ and the assumed $e_\mathrm{cut}$ converge, i.e., where the gray line crosses the horizontal line $e/e_\mathrm{cut}=1$. The value of $e_\mathrm{cut}$ where this occurs is our measurement of the shape of the stacked voids. The blue histogram in Figure~\ref{fig:shape_distr} shows the distribution of these values for the 1000 mocks. The mean void shape is close to unity and the $1\sigma$ scatter in the mocks is about 2.6\%. This result is exactly what we expect since we have assumed the correct cosmology for the mocks and should thus recover a spherical void shape in real space.

We next measure the shapes of stacked voids in the 1000 mock catalogs in redshift space, using the same cosmology. This result shows the effects of redshift space distortions on void shapes and it also provides relevant predictions that can be directly compared to measurements from the BOSS CMASS data. The bottom panel of Figure~\ref{fig:e_distr} shows the $e/e_\mathrm{cut}$ as a function of $e_\mathrm{cut}$ measurements and the red histogram in Figure~\ref{fig:shape_distr} presents the distribution of stacked void shapes for the 1000 mocks. These results reveal that redshift space distortions cause a slight squashing of voids along the line of sight, with a mean void axis ratio of 0.916. This result is opposite from what we expect from linear theory, where voids are expanding in comoving space and should thus be elongated in redshift space, and it demonstrates that the dynamics of galaxies within the region we consider are non-linear. Other recent studies using the same void-finding algorithm also find a squashing, although the magnitude of the effect is somewhat larger in those works \citep[e.g.,][]{Lavaux:2012,Sutter:2014a,Pisani:2015}. The $1\sigma$ scatter in the mock shapes in redshift space is about 2.6\%, which serves as our best estimate of the uncertainty in the the measured shape of stacked voids in the BOSS CMASS data. Finally, we measure the shape of stacked voids from the BOSS galaxy catalog, assuming the same fiducial cosmology as the mocks, and we obtain a void axis ratio of 0.929. This measurement is shown as a black vertical line in Figure~\ref{fig:shape_distr} and it is entirely consistent with the distribution from the mocks. The dashed red ellipse in Figure~\ref{fig:stacking_visual_single} has an ellipticity of 0.929 and thus illustrates the true shape of the stacked void seen in the figure.

\begin{figure}
	\includegraphics[scale=0.45]{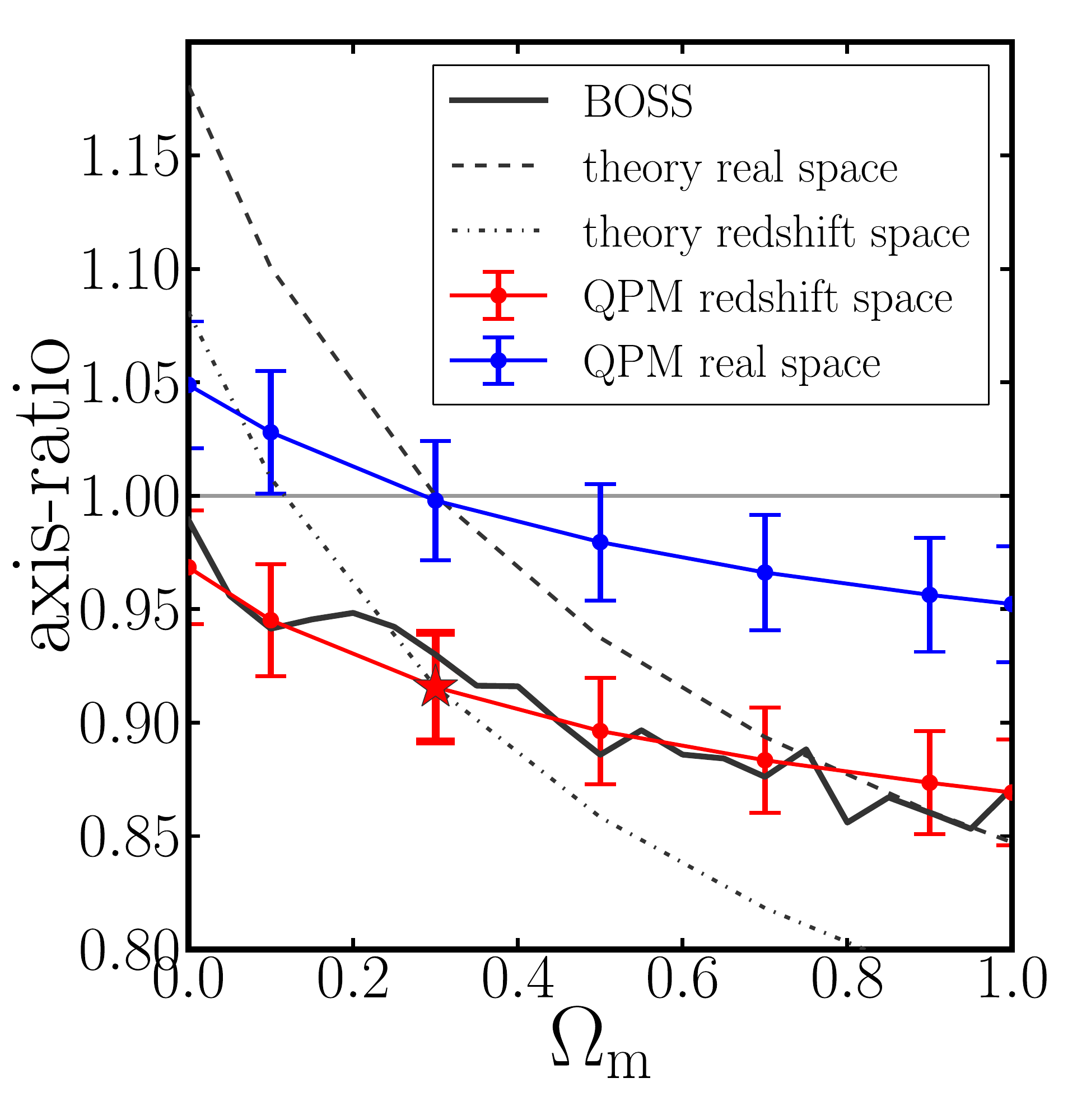}
	\caption{Shape measurements of the stacked voids assuming different values of $\Omegam$. Blue and red lines show the mean measurements from 1000 QPM mock catalogs in real space and redshift space, respectively, with error bars showing the standard deviation of the 1000 measurements. The black line shows the measurements from the BOSS CMASS galaxy catalog. The red point at $\Omegam=0.29$ is highlighted with a star to indicate that this is the shape of the stacked voids when assuming the correct cosmology and including redshift space distortion effects. The dashed line shows the ideal theoretical prediction in real space given by equation~\ref{eq:ratio}, and the dash-dotted line is the theoretical prediction in redshift space obtained by simply scaling the dashed line by the value of the red star point.}
	\label{fig:shape_vs_om}
\end{figure}

\section{Cosmological constraints}
\label{s:results}

To apply the AP test and obtain cosmological constraints from the BOSS stacked voids, we need to repeat our shape measurements for different assumed cosmologies. We assume a flat $\lcdm$ universe with a cosmological constant and we repeat the void shape measurements for a set of different $\Omegam$ values. For each $\Omegam$, we reconvert galaxy redshifts to line-of-sight comoving distances, re-run the void finding algorithm, stack the new set of voids, and measure the shape of the stacked voids using the methodology outlined in the previous section. We follow this procedure for both the QPM mocks and the BOSS data.

Figure~\ref{fig:shape_vs_om} shows the results of this procedure. The blue line shows the mean shape measurements of stacked voids from the 1000 QPM mocks in real space, as a function of the value of $\Omegam$ that was assumed. Error bars show the standard deviation of the 1000 measurements. The point at $\Omegam=0.29$ is simply the mean and standard deviation of the blue histogram in Figure~\ref{fig:shape_distr}. At this true mock cosmology, stacked voids are measured to be spherical, but this is not the case at other, incorrect, cosmologies. Values of $\Omegam$ that are too low or too high yield voids that are elongated or squashed along the line of sight, respectively. The black dashed line in Figure~\ref{fig:shape_vs_om} shows the theoretical AP prediction given by Equation~\ref{eq:ratio}, using the median BOSS CMASS redshift of 0.57. Ideally, the blue line should follow this dashed line. However, due to the sparse sampling of the galaxy distribution combined with the nature of the void finding algorithm, the measured shapes of stacked voids differ from the theoretical AP prediction. We discuss this issue more in \S\ref{s:discussion}.

The red line and errors bars in Figure~\ref{fig:shape_vs_om} show the axis ratio as a function of $\Omegam$ in redshift space. Redshift distortions cause an almost constant shift in void shape across the whole range of $\Omegam$. Voids in redshift space are on average $\sim8\%$ squashed along the line of sight. The point at $\Omegam=0.29$ is highlighted with a red star because this represents the predicted shape of BOSS stacked voids when we assume the correct cosmology. The dot-dashed line shows the result of multiplying the theoretical AP prediction (shown by the dashed line) by a factor 0.916. In other words, this is the predicted AP signal in redshift space if we simply apply a constant correction factor to account for redshift distortions. However, this is not correct since sparse sampling causes measured void shapes to be more spherical, i.e., the red line is flatter than the dot-dashed line. This feature is unfortunate because the flatter slope of the red line means larger uncertainty in $\Omegam$ constraints. Applying a simple correction for redshift distortions and using the analytic AP shape given by Equation~\ref{eq:ratio} is not correct and will cause cosmological parameter uncertainties to be underestimated. The solid black line in Figure~\ref{fig:shape_vs_om} indicates the measurement from the BOSS CMASS galaxy catalog, and it is consistent with the measurements of mocks in redshift space. The AP signal is clearly detected.

\begin{figure}
	\includegraphics[scale=0.45]{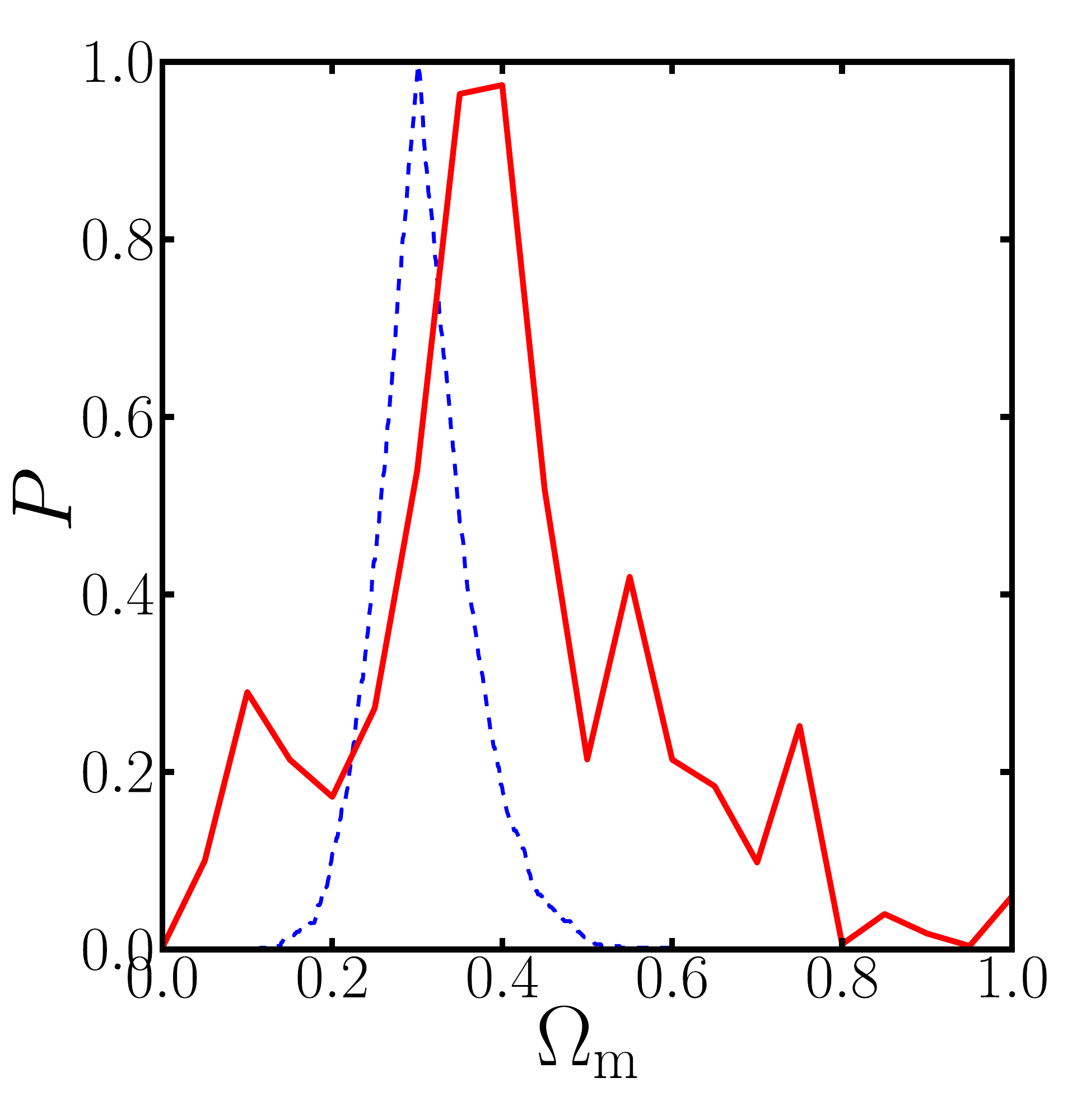}
	\caption{The probability distribution of $\Omegam$ (red curve) calculated by comparing the shape measurements of the CMASS data (shown by the black line in Fig.~\ref{fig:shape_vs_om}) to the expected distribution measured from the mock catalogs in redshift space and assuming the correct cosmological model (shown by the red histogram in Fig.~\ref{fig:shape_distr}). Specifically, the probability is given by the fraction of such mock catalogs that have a higher stacked void ellipticity than that measured from CMASS. The blue dashed curve indicates the optimal constraint by replacing the CMASS measurement with the theoretical AP prediction (dash-dotted curve in figure~\ref{fig:shape_vs_om}). The probability distribution of $\Omegam$ resulting from this analysis is not Gaussian, but has a narrow peaked shape.}
	\label{fig:om_probability}
\end{figure}

Mock catalogs that mimic the survey geometry and density are essential for deriving accurate cosmological constraints from the AP test on stacked voids. We take the distribution of stacked void shapes of the 1000 QPM mocks in redshift space for $\Omegam=0.29$ (the red star in Fig.~\ref{fig:shape_vs_om} or the red histogram in Fig.~\ref{fig:shape_distr}) to represent the probability distribution of measurements one might make for the BOSS CMASS sample if one assumes the correct cosmological model. We then compare the actual measurement from BOSS CMASS at each assumed value of $\Omegam$ (shown by the black line in Fig.~\ref{fig:shape_vs_om}) to this distribution to obtain a likelihood for that value. For example, when we assume $\Omegam=0.2$, we measure a shape of 0.948 for the BOSS CMASS stacked voids. Only 86 of the 1000 QPM mocks in redshift space and assuming the correct $\Omegam=0.29$ have shapes higher than this eccentricity. We therefore assign the value of $\Omegam=0.2$ to have a probability of 0.172 (two-tailed test). The red curve in Figure~\ref{fig:om_probability} shows the resulting probability distribution for $\Omegam$. Taking the mean and 68\% percentile confidence region of this distribution produces a measurement of $\Omegam = 0.38^{+0.18}_{-0.15}$.

As we discussed above, sparse sampling prevents us from obtaining the steep relationship between $\Omegam$ and void shape given by Equation~\ref{eq:ratio}. However, it is interesting to check what cosmological constraints are produced in this ideal case. To perform an estimate, we recalculate the probability distribution for $\Omegam$, but replacing the black line in Figure~\ref{fig:shape_vs_om} by the dot-dashed line in the same figure. The resulting distribution is represented by the dashed blue curve in Figure~\ref{fig:om_probability}. In this ideal case, the error in $\Omegam$ drops to $\sim0.05$, which is more than three times smaller than the value reported in this paper. The probability distribution for $\Omegam$ is quite different from Gaussian and has a narrow peaked shape. If the shape distribution predicted from mocks (the red histogram in Fig.~\ref{fig:shape_distr}) is Gaussian, which is close to correct, then the probability corresponding to any observed value $e$ is simply the integral of this Gaussian from zero to $e$ or from $e$ to infinity, depending on whether $e$ is smaller or greater than the mean mock shape, times two for the two-tailed test. If $e(\Omegam)$ were a linear function, then the resulting probability distribution of $\Omegam$ would simply be a symmetric version of the \texttt{erf} function. Instead, $e(\Omegam)$ has some curvature and becomes shallower with $\Omegam$, so the probability distribution has a longer tail at high values of $\Omegam$.

\section{Discussion}
\label{s:discussion}

In this study we have identified cosmic voids in the completed BOSS CMASS galaxy catalog and a set of 1000 QPM mock galaxy catalogs with the ZOBOV void finding algorithm, and we have accurately measured the shape of the stacked voids. By repeating this measurement for different assumed cosmological models, we have applied the Alcock-Paczy\'nski test on the stacked voids and placed constraints on the parameter $\Omegam$. 

\subsection{Redshift Space Distortions}

Redshift space distortions are the biggest concern in applying the AP test. This analysis finds that the shape of the stacked voids in redshift space is squashed along the line of sight by a factor of 0.92. While one might expect the stacked voids to be elongated in redshift space, as predicted by linear theory, the opposite is true. Other recent studies using the ZOBOV void finding algorithm also reveal void squashing. For example, \citet{Lavaux:2012} found that the squashing effect is almost universal and constant, though they only used high density N-body simulation particles as the tracers and the voids they studied were much smaller. \citet{Sutter:2014a} used more realistic mock catalogs to demonstrate that the squashing appears universal for all void sizes, at all redshifts, and for all tracer densities. However, these studies reported squashing at the 14\% level, which is twice as large as the effect we find here. It is difficult to compare our results directly because the galaxy samples are different, the ZOBOV options are different, and the details of the mock catalogs used are different. In particular, allowing voids with high density minima, as \citet{Sutter:2014a} do, produces more voids in overdense (and therefore collapsing) regions \citep{Nadathur:2015}. In addition, \citet{Pisani:2015} showed that peculiar velocities affect small voids more than large voids so the difference could be in part due to different size distributions of voids. It is important to study this issue further. To deal with redshift distortions, previous AP studies \citep{Sutter:2012b,Sutter:2014a} corrected their shape measurements empirically to recover the real space shape of stacked voids. We do not apply a correction, but we compare our measurement directly to mock catalogs in redshift space, which is essentially the same thing. 

The main danger in this empirical approach is that we have only measured the effect of redshift distortions for a single fiducial cosmological model (the model used to construct the mock catalogs) and we are implicitly assuming that it does not depend on cosmology as we vary $\Omegam$ in the AP test. This cannot be true in detail since peculiar velocities depend on $\Omegam$. We probe this issue by running a set of N-body simulations for different values of $\Omegam$ spanning the range 0.2--0.4, and constructing mock catalogs with the same number density of galaxies as in the BOSS CMASS sample. These mocks have a volume of $1\hgvol$, which is smaller than that of our BOSS CMASS sample, but they are large enough to detect a sizeable effect. We examine the ratio of stacked void shapes in redshift space to real space and find no trend with $\Omegam$. This result indicates that variations in the effect of redshift distortions due to cosmology must be weak and are overwhelmed by the shape noise in the stacked voids. It would be better to repeat this test with much larger mock catalogs in order to accurately measure the cosmological dependence of redshift distortions and thus include this dependence in the AP analysis. This effort is beyond the scope of this current study, but will become necessary as future survey volumes grow larger and statistical errors tighten.

\subsection{Other Statistical and Systematic Errors}

Our tests with mock catalogs demonstrate that we measure the shape of the stacked voids with a precision of $2.6\%$ (see Fig.~\ref{fig:shape_distr}). To understand how much of this statistical error arises from Poisson noise, we generate a large number of realizations of mock spheres filled with random points with the same number density and a similar density profile as the stacked voids from the CMASS sample, and we measure the dispersion in their shapes. We find that Poisson noise only contributes approximately $0.6\%$ to the statistical error. The dominant error is due to the limited number of voids in the survey volume combined with the variance in the shapes of individual voids, i.e., shape noise. As a result, to reduce the statistical errors in the stacked void shape measurement, it is more important to increase the volume of the survey (which will lead to more voids) than to increase the density of tracers.

We have shown in this work that it is not possible to recover the theoretical AP prediction for how the stacked void shape depends on cosmology. This conclusion is due to the impact of sparse sampling of the density field, which causes all measured void shapes to be more spherical, as reported by \citet{Sutter:2014b}. Sparse sampling prevents the ZOBOV algorithm from perfectly finding all the voids in their natural shape and extension. For example, in a sparse sample, an elongated void might be identified as two separate voids, neither of which is as elongated as the original void. Another way to understand this is that the idealized AP signal applies to the case of structures whose intrinsic shapes do not depend on cosmology. One could then observe them deform under the coordinate transformation that occurs with changing cosmology. However, ZOBOV identifies voids using galaxies whose redshift-space positions depend on the assumed cosmology and it will thus select different structures as the cosmology changes. In the extreme case of a Poisson distribution of points, ZOBOV will find voids, but their average shape will always be spherical because random fluctuations do not have preferred orientations. This shape does not depend on the assumed cosmology and if one changed the cosmology, ZOBOV would just identify a different set of voids. The real case is somewhere between the ideal case and the extreme Poisson case: voids are real structures, but sparse sampling makes them behave a bit like a Poisson distribution.

Unfortunately, the result of this effect is to significantly degrade the constraining power of the stacked void AP test (see Fig.~\ref{fig:om_probability}). Moreover, this effect mandates that we use mock catalogs to calibrate the AP effect, since the analytic expression given by Equation~\ref{eq:ratio} does not apply to samples of realistic density. We explore to what extent sparse sampling affects our results using a set of dark matter halo catalogs from the LasDamas project \citep{McBride:2009}. We create halo catalogs with different halo mass cuts to generate samples with different sparseness. We then stretch the catalogs in the line of sight direction and we run ZOBOV followed by our stacking and shape measurement method on all the halo catalogs. We find that the denser the sample, the closer we can recover the actual input stretch. The implication of this is that a future survey with higher galaxy number density will provide a stronger constraint not only by slightly improving the Poisson statistical error, but also by bringing the shape measurements closer to the theoretical AP prediction. Unfortunately, our tests reveal that this improvement is likely to be modest, even for a galaxy density that is 15 times higher than BOSS CMASS. A more fruitful avenue for improving constraints may be to search for alternate void finding algorithms that are less sensitive to galaxy sparseness.

\begin{figure}
	\includegraphics[scale=0.44]{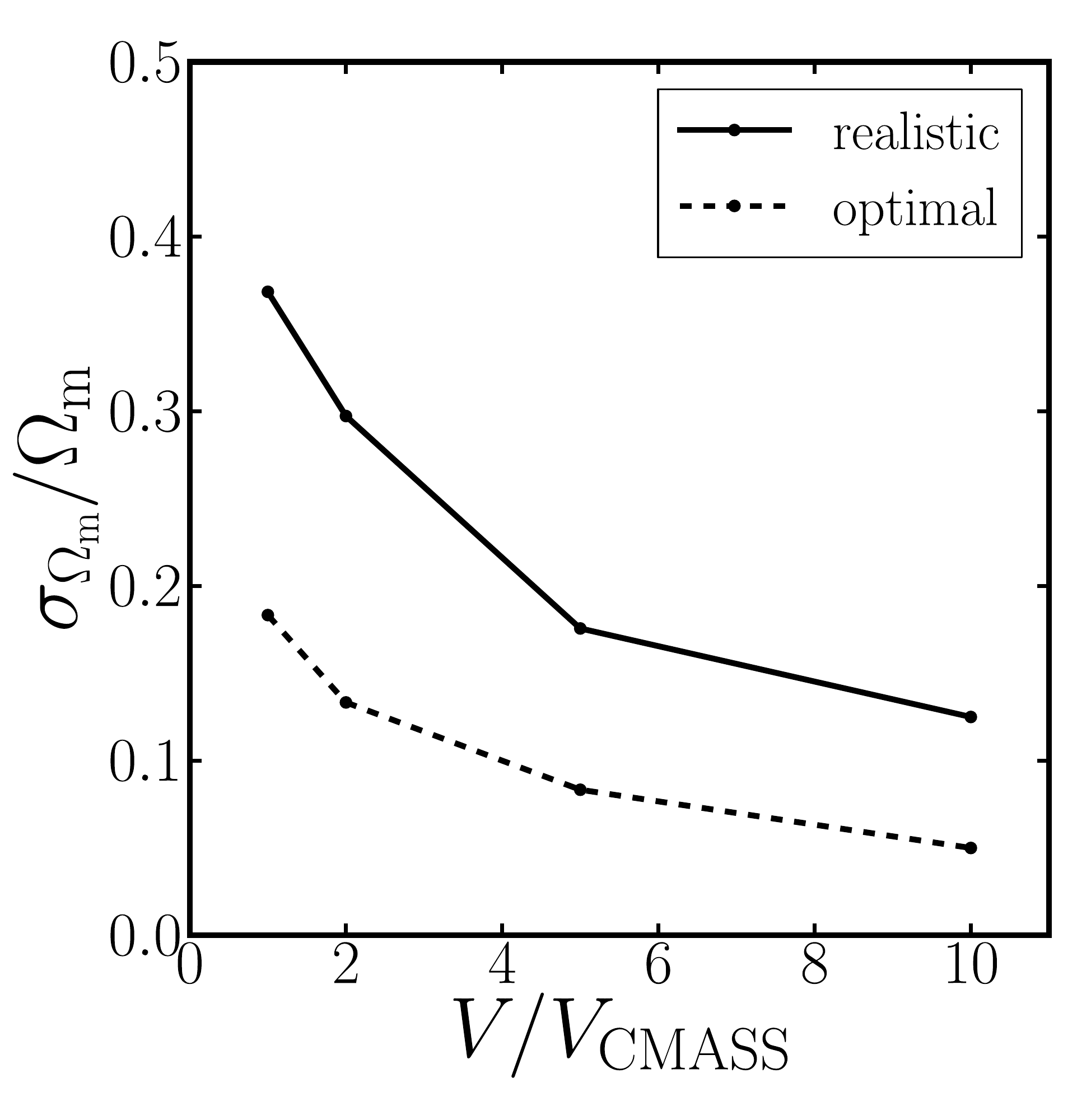}
	\caption{Projections of the uncertainty in $\Omegam$ as a function of survey volume in units of the BOSS CMASS volume. Projections are made by combining multiple CMASS mocks to mimic a larger survey volume and measuring the corresponding distribution of measured stacked void shapes. The ``realistic" case (solid curve) assumes that our ability to measure the true AP signal will remain the same as it is in this study, i.e., we use a polynomial fit to the black line in Fig.~\ref{fig:shape_vs_om} to calculate the probability distribution of $\Omegam$. The ``optimal" case assumes that we will be able to perfectly recover the true AP signal and so we use the dot-dashed line in Figure~\ref{fig:shape_vs_om} instead.}
	\label{fig:prediction_err}
\end{figure}

\subsection{Forecast for Future Surveys}

We now attempt to make realistic projections of the uncertainty in $\Omegam$ for future surveys. To estimate how the uncertainty scales with survey volume, we stack together the voids from multiple mock catalogs (from our set of 1000), in order to mimic a larger survey volume. For example, we combine sets of five mocks to mimic a survey that is five times larger than the BOSS CMASS sample, which produces 200 such mock samples in total. We then calculate the standard deviation of shape measurements from these new samples and assume a Gaussian shape distribution to recalculate the probability distribution of $\Omegam$. Figure~\ref{fig:prediction_err} displays the resulting fractional error in $\Omegam$ as a function of survey volume in units of the BOSS CMASS volume (which is $3.7\hgvol$). We present two cases of error projections. The ``realistic" case assumes that our ability to measure the true AP signal will remain the same as it is in this study. In other words, we use the black line in Figure~\ref{fig:shape_vs_om} (actually, we use a polynomial fit to this line) to calculate the probability distribution of $\Omegam$. The ``optimal" case assumes that we will be able to perfectly recover the true AP signal and so we use the dot-dashed line in Figure~\ref{fig:shape_vs_om} instead. The uncertainty in $\Omegam$ roughly scales with square root of the survey volume. For a future survey of 10 times the BOSS CMASS sample, we can expect to measure $\Omegam$ with 12\% accuracy using the same technique and procedure presented in this paper. In the optimal case where we can perfectly retrieve the natural extension of every void and fully recover the AP signal, we can expect an accuracy of 5\% via the AP test on the stacked voids. This prediction assumes larger surveys with the same number density and redshift range as the BOSS survey. Future galaxy redshift surveys will extend to higher redshifts and attain higher densities, which will both benefit the AP test. 

The cosmological constraints obtained in this paper using the final BOSS data are substantially weaker than those forecast by \citet{Lavaux:2012}, who predicted that the void AP test applied to BOSS would rival constraints using the BAO method. This discrepancy is due to several factors. First, \citet{Lavaux:2012} used voids identified in the dark matter distribution of N-body 
simulations, which yielded a large number of fairly small voids. In the actual sparse BOSS data, we can only identify a relatively small number of much larger voids. Second, the small number of voids combined with the fairly narrow redshift range of the BOSS CMASS sample do not allow an investigation of the AP signal as a function of redshift, which would add constraining power. Third, we have shown that the sparseness of the BOSS data weakens the AP signal relative to the theoretical expectation by a factor of $\sim 2-3$. Finally, \citet{Lavaux:2012} used a different methodology for estimating the ellipticity of stacked voids that relies on fitting a model to the stacked density profile. This method does not work well when the total number of voids in a stack is small \citep{Sutter:2014a}.

As we were preparing to submit this paper, an analysis appeared on the arXiv by \citet{Hamaus:2016} who use voids in the BOSS DR11 data to obtain cosmological constraints that are $\sim4$ times stronger than what we present here (comparing the fractional error in $\Omegam$). There are several factors that may be partly responsible for this. First, \citet{Hamaus:2016} have a much larger sample of voids because they do not make the cuts we adopt to maximize the physicality of our void sample. Specifically, they do not impose a cut to ensure that their voids are statistically significant underdense regions, whereas we only use voids that have a higher significance than $2\sigma$ compared to a Poisson sample. However, they do appear to find a signal from even the low-significance voids, suggesting that stacking enough low-significance voids may be justified. Second, the \citet{Hamaus:2016} analysis uses a physical model of the peculiar velocities around voids, which contain much of the cosmological signal. Their method is thus a hybrid RSD/AP method, rather than the pure AP method. Third, \citet{Hamaus:2016} measure and model the void-galaxy cross-correlation function at all radii, rather than measure a single ellipticity for each void stack. Finally, they use the analytic formula (Eq.~\ref{eq:ratio}) for the AP effect, which we demonstrated for our methodology underestimates the cosmology error bars due to discreteness (see end of \S~\ref{s:results}). This problem may apply to their methodology as well. If this is the case, reliably estimating their cosmological error bars will require them to compare to mock catalogs that are analyzed using a grid of assumed cosmological models, as we have done in this study. It remains to be seen in what circumstances the full sensitivity of the analytic formula can be realized. It is important to study all of these differences in more detail.

There is much room for further study and improvement to the methods used here. Most importantly, the effects of redshift distortions must be better understood and quantified as a function of cosmology. It would also be valuable to explore modifications to the void finding algorithm in order to find a method that is more robust against the sparseness of the galaxy distribution. It is worth investigating binning voids into different size and redshift bins with the goal of optimizing their constraining power. Overall, the optimal strategy would likely be to investigate the signal-to-noise ratio as a function of both void parameters and galaxy location within the void, and weight accordingly. Future galaxy redshift surveys, such as eBOSS \citep{Dawson:2016}, DESI \citep{Levi:2013}, Euclid \citep{Laureijs:2011} and WFIRST \citep{Spergel:2013}, will produce even larger maps of galaxies in the next decade, which can potentially make cosmic voids a very powerful tool to constrain cosmology. However, whether the AP effect using voids will be competitive with other cosmological probes remains to be determined.

\acknowledgments

We thank Paul Sutter, Jennifer Piscionere, Manodeep Sinha, David Weinberg, Nico Hamaus, Alice Pisani, and Ben Wandelt for valuable discussions. Q.M. and A.A.B. were supported in part by the National Science Foundation (NSF) through NSF Career Award AST-1151650. R.J.S. is supported by DOE grant DE-SC0011981. We thank the Advanced Computing Center for Research and Education (ACCRE) at Vanderbilt for providing computational resources that were used in this work. This work also used the Extreme Science and Engineering Discovery Environment (XSEDE), which is supported by National Science Foundation grant number ACI-1053575. In particular, we used the high performance computing and storage resources at the Texas Advanced Computing Center (TACC).

Funding for SDSS-III has been provided by the Alfred P. Sloan Foundation, the Participating Institutions, the National Science Foundation, and the U.S. Department of Energy Office of Science. The SDSS-III web site is http://www.sdss3.org/.

SDSS-III is managed by the Astrophysical Research Consortium for the Participating Institutions of the SDSS-III Collaboration including the University of Arizona, the Brazilian Participation Group, Brookhaven National Laboratory, Carnegie Mellon University, University of Florida, the French Participation Group, the German Participation Group, Harvard University, the Instituto de Astrofisica de Canarias, the Michigan State/Notre Dame/JINA Participation Group, Johns Hopkins University, Lawrence Berkeley National Laboratory, Max Planck Institute for Astrophysics, Max Planck Institute for Extraterrestrial Physics, New Mexico State University, New York University, Ohio State University, Pennsylvania State University, University of Portsmouth, Princeton University, the Spanish Participation Group, University of Tokyo, University of Utah, Vanderbilt University, University of Virginia, University of Washington, and Yale University.

\bibliographystyle{apj}
\bibliography{void_ap}

\end{document}